\documentclass[journal]{IEEEtran}

\usepackage[cmex10]{amsmath}
\usepackage{enumitem}
\usepackage{amsfonts}
\usepackage{booktabs}
\usepackage[font={small}]{caption}
\usepackage{subfig}
\usepackage{relsize}
\usepackage{hyperref}
\usepackage{graphicx}
\usepackage{algorithm}
\usepackage{algpseudocode}
\usepackage{steinmetz}

\newtheorem{theorem}{\textbf{Theorem}}
\newtheorem{corollary}{\textbf{Corollary}}[theorem]

\makeatletter
\def\BState{\State\hskip-\ALG@thistlm}
\makeatother

\begin{document}

\title{On the Exploitation of Admittance Measurements for Wired Network Topology Derivation}

\author{\IEEEauthorblockN{Federico Passerini and Andrea M. Tonello}\\
\IEEEauthorblockA{EcoSys Lab,
University of Klagenfurt, Austria\\ Email: {\{federico.passerini, andrea.tonello\}@aau.at}}}

\maketitle

\begin{abstract}
The knowledge of the topology of a wired network is often of fundamental importance. For instance, in the context of  Power Line Communications (PLC) networks it is helpful to implement data routing strategies, while in power distribution networks and Smart Micro Grids (SMG) it is required for grid monitoring and for power flow management. In this paper, we use the transmission line theory to shed new light and to show how the topological properties of a wired network can be found exploiting admittance measurements at the nodes. An analytic proof is reported to show that the derivation of the topology can be done in complex networks under certain assumptions. We also analyze the effect of the network background noise on admittance measurements. In this respect, we propose a topology derivation algorithm that works in the presence of noise. We finally analyze the performance of the algorithm using values that are typical of power line distribution networks.
\end{abstract}



%
\IEEEpeerreviewmaketitle

\section{Introduction}

\IEEEPARstart{W}{ired} data transmission networks like telephone and digital subscriber line (DSL) networks are well established communication technologies that allow the exchange of enormous amount of data among all the users therein connected. In recent years also power lines have been extensively investigated for data transmission, enabling any user connected to a power grid to exchange information \cite{lampe2016power}. This new technology contributed in fostering the advent of smart grids \cite{gungor2011sgt}: not only mere infrastructures to distribute energy to users, but also intelligent networks that exchange information in order to efficiently satisfy the user demands and manage bidirectional power flows.

In this paper, we address the problem of identifying the topology of such wired data transmission networks, considering but not restricting to PLC in the context of distribution grids and SMG as a possible application. The term topology herein refers to the network graph that describes the nodes relative displacement, and the length of the wired connections.

The identification of the network topology is important in many respects. In the context of DSL networks, it is part of the line qualification procedures that are used to assess the ability of a specific wired network to support different DSL services before the actual deployment. A proper knowledge of the network topology allows to compute the channel transfer functions and can also be used for support engineering and maintenance operations \cite{1007375,5256192,5953512}.
In the context of SMG, not only communication is involved, but also power transmission. PLC are used herein as a mean of controlling and monitoring of the grid. The knowledge of the network topology is a fundamental requirement to develop routing strategies for both power and data information, as well as coordination algorithms for distributed computation \cite{1430477}.  

Some recent proposals \cite{ahmed2012topology2,7431885,erseghe2013topology,lampe2013tomography} aim to estimate the PLC network topology as plug-and-play solutions (i.e., no historical data is considered), by using a two-step procedure. First, the channel response is sensed at different frequencies to estimate the distances between the nodes. Subsequently, different algorithms are applied to infer the network topology. This two-step procedure can be repeated over time, so that the topology is updated when the state of the network changes. In \cite{ahmed2012topology2} Frequency Domain Reflectometry (FDR) is used to perform a single-end distance measurement between one node and all the others. However, the FDR reliability is limited by the maximum observable distance and the number of branches. Some better performance is achieved using Time Frequency Domain Reflectometry \cite{7431885}. In \cite{erseghe2013topology}, it is assumed that all the nodes of the network are equipped with a PLC modem. The distance between nodes is estimated via Time of Arrival (ToA), i.e., from the propagation delay of the transmitted signal using an energy detector. ToA estimation deploying PLC modems is also used in \cite{lampe2013tomography}, where the energy detector is compared to a sub-space estimation method that is generally more efficient. 

In this paper, we present a novel technique for wired network topology estimation that does not rely on historical data and uses admittance measurements operated at all the network nodes. This last requirement is envisioned for a future SMG scenario in which all the nodes will be equipped with PLC modems \cite{erseghe2013topology}. To our knowledge, admittance measurements have already been used in power networks in order to implement efficient fault detection strategies \cite{6338332},\cite{6954542}, but never for topology estimation. On the other hand, equivalent S-parameter and impedance measurement have been used in DSL to identify the DSL topology \cite{5256192} and the channel transfer function \cite{5953512} under ideal conditions. The purpose of this paper is to discuss the role of admittance measurements, the fundamental aspects, the conditions and the assumptions that have to be made to allow the derivation of a wireline network topology via admittance measurements. The focal points are the application of the Transmission Line (TL) theory \cite{Pozar} and the measurement of the network admittances at all nodes of the network. Based on this, we formulate an analytical result and a related algorithm which allow us to derive the topology exactly and independently from the size and complexity of the network, when no noise or measurement error is present. Hence, in this paper we use the terminology topology derivation instead of identification or estimation. An analytical formulation of the line background noise effect on the admittance measurement is then derived. Furthermore, an approach to derive the topology in the presence of noisy measurements is described, and its performance is assessed. 

\subsection{Relation with existing solutions and contribution}
The main aim of this paper is to introduce and thoroughly explain the theoretical aspects of a novel technique for topology derivation in wired networks, as an alternative or complement to the present techniques. Moreover, a section is devoted to expose the open issues and the possible directions that can be pursued to improve the topology derivation method presented in this paper, and the topology inference methods in general.

The proposed technique can tackle some limitations of the existing plug-and-play topology estimation techniques for PLC networks, namely \cite{ahmed2012topology2,7431885,erseghe2013topology,lampe2013tomography}, but at the same time introduces some challenges, as discussed below.

	\emph{Meter}: the existing techniques require PLC modems or reflectometers. Our approach relies on the use of just impedance or voltage meters, which can also be embedded in PLC modems.
	
	\emph{Channel model}: the existing techniques use a phenomenological channel model that requires some assumptions about the channel (propagation velocity equal for each cable, constant reflection coefficients over frequency, propagation constant be linear function of frequency). If some of the assumptions do not hold true, this might deteriorate the accuracy of the topology estimation. Our technique uses a physical channel model based on TL theory \cite{tonello2011bottomup}, which is more strictly related to the physics of propagation. On the other hand it requires information about all the loads and cable parameters, which might not always be available.
	
	\emph{Operating frequency}: High frequencies and large bandwidths are needed to obtain reasonable performance of ToA techniques and good resolution in FDR or TFDR. Our approach operates at a single frequency that can also be in the range of the narrow band PLC spectrum, more commonly used in SMG.
	
	\emph{Dimension of the network}: a problem of the existing two step techniques \cite{ahmed2012topology2,7431885,erseghe2013topology,lampe2013tomography}, is that the topology is inferred only if each node (or the main node in the case of FDR and TFDR) knows the distance between itself and any other node of the network. Moreover, the multipath propagation and the strong attenuation of the high frequency signals limit the maximum distance that can be sensed. This problem can be solved by splitting the network in many overlapped subsections (see \cite{lampe2013tomography2}). In our approach each node finds the distance only to its neighbor and can contextually infer its topological position. Hence, our approach is not limited by the dimension of the network, but by the maximum distance between two neighbor nodes.

At the same time our work shares some similarities with \cite{5256192}. Both of them use admittance or equivalent scattering parameter measurements and assume all the line and load parameters to be ideal. They however differ on the requirements, the algorithms and the final results. The work in \cite{5256192} applies a genetic algorithm to derive the network topology and the number of nodes starting from single or double end measurements. However the algorithm is tested on simple networks, and no parameter error or noise is taken into account. The work presented in this paper relies on measurements performed at every node of the network, but it can derive the topology for any network, when no noise is considered. Furthermore, an analysis of the impact of network noise on the derivation algorithm is performed.

The reminder of this paper is organized as follows. In Section~\ref{sec:ana}, a brief review of the basic equations of the TL theory used for distance computation is given. In Section~\ref{sec:sc2}, the main system of equations to solve the topology derivation problem is derived. In Section \ref{sec:noise}, the influence of the line network noise on admittance measurements is discussed. The topology derivation algorithm is then presented in Section~\ref{sec:topdev}. Numerical results are also reported in Section~\ref{sec:results} to study the robustness of the proposed algorithm to noise. Further remarks and open problems are also discussed in Section~\ref{sec:remarks}. Finally, the conclusion follows.

\section{Analytical evidence from TL theory}
\label{sec:ana}

In this section, we examine the TL theory in order to derive an equation that relates the length of the line that connects two nodes with the admittance at one end. We start by considering the simplest case of an unbranched transmission line of length $d$ that connects a Thevenin generator to a load $L$. The network admittance $Y(d)$ seen by the generator can be written as \cite{Pozar}
\begin{equation}
	Y(d) = Y_C \frac{1-\rho_L e^{-2\Gamma d}}{1+\rho_L e^{-2\Gamma d}},
	\label{eq:adm_TL}
\end{equation}
We refer to this relation as \emph{carry-back} equation since the load admittance is carried back to the input of the line to obtain $Y(d)$.  
$Y_C$ is the characteristic admittance of the line and $\Gamma = \alpha + j\beta $ is the propagation constant of the line, where $\beta = 2\pi/\lambda$ and $\lambda$ is the wavelength used to perform the measurement. $\rho_L$ is the load reflection coefficient written as
\begin{equation}
	\rho_L = \frac{Y_C - Y_L}{Y_C + Y_L},
	\label{eq:rho_TL}
\end{equation}
where $Y_L$ is the load admittance. All the aforementioned quantities (except $d$) depend on the frequency. Herein and in the following, this dependency is implicit to ease the notation. However, as it will be discussed, the choice of the frequency influences the algorithm.  

From \eqref{eq:adm_TL}, under the assumption that we know the load reflection coefficient \eqref{eq:rho_TL}, an equation that relate $d$ to the measured network admittance can be found\footnote{Actually the equations are two. The second one is
$
	d = \frac{1}{2\beta}\left(\phase{\rho(d)}-\phase{\rho_L}\right)
$,
but it limits the estimation of $d$ due to the intrinsic periodicity of the phase.}:
\begin{equation}
	d = \frac{1}{2\alpha}\left(\log{\left|\rho(d)\right|}-\log{\left|\rho_L\right|}\right)
	\label{eq:l_TL}
\end{equation}
where 
\begin{equation}
	\rho(d) = (Y_C - Y(d))/(Y_C + Y(d))
	\label{eq:rho_simple}
\end{equation} 
and $k$ is an integer number. \eqref{eq:l_TL} is bijective since in real lines $\alpha \ne 0$, so that $\left|\rho(d)\right|$ is a monotonically decreasing function of $d$. We also remark that in two cases it is impossible to find the length of the transmission line using \eqref{eq:l_TL}:
\begin{enumerate}
	\item if the transmission line is ideal, i.e. $\alpha = 0$. In fact in this case $Y(d)$ would be a periodic function of $d$ with period $\lambda/2$.
	\item if $\rho_L=0$, i.e. when the load is matched to the transmission line impedance. In fact in this case $Y(d)$ would simply be a constant.
\end{enumerate}

When branches attached to the main transmission line are considered, the problem of finding the length of each line becomes more difficult. In the following sections, we consider a complex network made of $N$ nodes. We refer to the nodes as branch nodes (identified by line intersections) and termination nodes (leaves). The known parameters in such a network are the cable parameters $\Gamma_i$, $Y_{C_i}$ and the loads $Y_{L_i}$ or equivalently the reflection coefficients $\rho_{L_i}$ $\forall i \in [1,\dots N]$, and network admittances $Y_{i}$ at every node of the network.


\section{Derivation of lines' length with unknown topology graph}
\label{sec:sc2}

In this section, we derive the analytical formulas and the theorem that allow us to derive the network topology.

\begin{figure}
	\centering
		\includegraphics[width=0.30\textwidth]{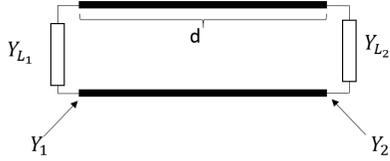}
	\caption{Sketch of a simple 2-loads network}
	\label{fig:two_loads}
\end{figure}

\begin{figure}
	\centering
		\includegraphics[width=0.4\textwidth]{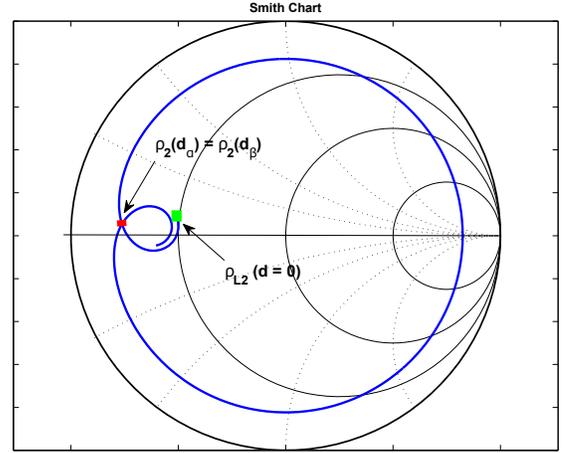}
	\caption{Smith chart with a possible realization of $\rho_2$ according to equation \eqref{eq:rhomale}.}
	\label{fig:rho_example}
\end{figure}

To understand how to proceed, we initially consider the simple network depicted in Fig. \ref{fig:two_loads}.
Herein, two loads are connected by a line with length $d$, and the network admittances $Y_1$ and $Y_2$ are measured at the line ends. Using \eqref{eq:adm_TL} and \eqref{eq:rho_TL} it is possible to relate the network admittance $Y_1$ as a function of $d$ as follows:
\begin{equation}
	\begin{cases}
		\mathlarger{Y_1 = Y_C \frac{1-\rho_{L_2} e^{-2\Gamma d}}{1+\rho_{L_2} e^{-2\Gamma d}} + Y_{L_1}}\\[10pt]
		\mathlarger{\rho_{L_2} = \frac{Y_C - Y_{L_2}}{Y_C + Y_{L_2}} = \frac{Y_C-Y_2+Y_C \frac{1-\rho_{L_1} e^{-2\Gamma d}}{1+\rho_{L_1} e^{-2													\Gamma d}}}{Y_C+Y_2-Y_C \frac{1-\rho_{L_1} e^{-2\Gamma d}}{1+\rho_{L_1} e^{-2\Gamma d}}}}
	\end{cases}
	\label{eq:two_loads}
\end{equation}
A similar system of equations can be written for the network admittance $Y_2$ measured at node $2$. In this system, exploiting the second relation, we can write that  
\begin{equation}
	Y_1 = f(Y_2,Y_{L_1},Y_C,\Gamma,d), 
	\label{eq:fund}
\end{equation}	
so that the admittance measured at node $1$ is a function, in particular, of the network admittance $Y_2$ while the knowledge of the load admittance $Y_{L_2}$ is not explicitly required. It is important to point out that in general there can be two values of $d$ that are admissible, i.e., two possible solutions. This is because the term (load reflection coefficient) 
\begin{equation}
\rho_2 = \rho_{L_2}e^{-2\Gamma d}
\label{eq:rhomale}
\end{equation} 
that appears in the first equation of \eqref{eq:two_loads} may be such that 
\begin{equation*}
\rho_2(d_\alpha) = \rho_2(d_\beta)
\end{equation*}
for  $d_\alpha < \lambda/4$ and $\lambda/4 < d_\beta < \lambda/2$, depending on the parameters; 
in fact, $\rho_2$ is the product of an exponential function with $\rho_{L_2}$, which also has an exponential trend. Using the Smith chart, an example of $\rho_2$ as a function of $d$ is plotted in Fig. \ref{fig:rho_example}. It can therein be clearly seen that there are two distances in correspondence of which the reflection coefficient takes the same value. Given this fact, $\rho_2$ is not a bijective function of $d$, conversely from \eqref{eq:rho_simple} that exploits the load admittance $Y_{L_2}$ instead of the network admittance $Y_2$. It follows that to grant a unique solution for the variable $d$, we must assume that measurements have to be taken at a wavelength $\lambda \geq 4d$. 

\begin{figure}
	\centering
		\includegraphics[width=0.40\textwidth]{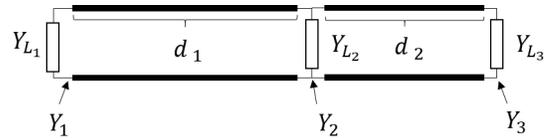}
	\caption{Sketch of a simple 3-loads network}
	\label{fig:three_loads}
\end{figure}

When the results shown above are extended to a more complex network, as for example the one depicted in Fig. \ref{fig:three_loads},
it is possible to write a system of equations similar to \eqref{eq:two_loads}. Now, the second equation of the system changes into
\begin{equation}
	\rho_{L_2} = \frac{Y_C - Y_{L2eq}}{Y_C + Y_{L2eq}} = \frac{Y_C - (Y_{L_2}+Y_{cb_{3}})}{Y_C + (Y_{L_2}+Y_{cb_{3}})}
	\label{eq:rho_three}
\end{equation}
where $Y_{cb_{3}}$ represents the load admittance $Y_{L_3}$ carried back to node 2. One can also write
\begin{equation}
	Y_{L_2}= Y_2 - Y_C \left(\frac{1-\rho_{L_1} e^{-2\Gamma_1 d_1}}{1+\rho_{L_1} e^{-2\Gamma_1 d_1}} + 
						\frac{1-\rho_{L_3} e^{-2\Gamma_2 d_2}}{1+\rho_{L_3} e^{-2\Gamma_2 d_2}}\right)
	\label{eq:Y2}
\end{equation}
so that with some simple algebraic manipulation, \eqref{eq:rho_three} finally becomes
\begin{equation}
	\rho_{L_2} = \frac{Y_C-Y_2+Y_C \frac{1-\rho_{L_1} e^{-2\Gamma_1 d_1}}{1+\rho_{L_1} e^{-2\Gamma_1 d_1}}}
							{Y_C+Y_2-Y_C \frac{1-\rho_{L_1} e^{-2\Gamma_1 d_1}}{1+\rho_{L_1} e^{-2\Gamma_1 d_1}}}
	\label{eq:rho2_long}
\end{equation}
that is equal to the second equation in the system \eqref{eq:two_loads}. Hence, this proves that equation \eqref{eq:fund} is still valid when another part of the network is branched to node 2 so that to find the distance $d_1$ we need a network admittance measurement at nodes $1$ and $2$ and apply the first equation in \eqref{eq:two_loads}. Then, to obtain $d_2$ we can proceed with a similar reasoning so that we relate according to a similar set of equations  $Y_2$ with $Y_3$. 

Furthermore, the system \eqref{eq:two_loads} of equations can be written as a single complex equation in the unknown $d$. The resulting equation has the following quadratic form 
\begin{equation*}
	\gamma e^{-2\Gamma d} + \delta e^{-4\Gamma d} + \epsilon = 0,
\end{equation*}
where $\gamma$, $\delta$ and $\epsilon$ are polynomial functions of $Y_1$, $Y_2$, $Y_C$ and $Y_{L_1}$. The solutions can be written in closed form as
\begin{subequations}
	\begin{align}
		d_1 = \frac{1}{2\Gamma}\log{\left[-\frac{a\left(\sqrt{-b(cY^2_2+lY_2+e)}+fY_2+g\right)}{h}\right]}\\[20pt]
		d_2 = \frac{1}{2\Gamma}\log{\left[-\frac{a\left(-\sqrt{-b(cY^2_2+lY_2+e)}+fY_2+g\right)}{h}\right]}
	\end{align}
	\label{eq:dist}
\end{subequations}
where $a$, $b$, $c$, $e$, $f$, $g$, $h$, $k$ and $l$ are polynomial functions of $Y_1$, $Y_C$, $Y_{L_1}$ and $\Gamma$.  We wrote \eqref{eq:dist} as an explicit function of the sole $Y_2$, which is the network admittance of node 2 that is adjacent to node 1 with measured network admittance $Y_1$.

It should be observed that the admittance measured at one termination node (leaf) of the PL network depends only on the physical parameters of the cable to whom it is branched, on the length of the cable connecting it to the nearest node, and on the network admittance measured at this second node. However, although we have measured all the network admittances, we still do not know the topology graph and therefore we do not know the association between the admittances, i.e., we do not know what nodes/admittances are adjacent and what nodes are not directly connected by a line. It may be believed, at a first glance, that the system of equations \eqref{eq:two_loads}, and therefore \eqref{eq:dist}, applies to any pair of admittances measured in the network so that we always get a physically meaningful (although wrong) distance. In reality, since all the terms inside the logarithm in \eqref{eq:dist} are complex, $d_1$ and $d_2$ can be complex depending on $Y_2$. Of course the distance we are looking for must be a real number and this is the case when $Y_2$ is the true admittance of a node that is adjacent to the node 1 with admittance $Y_1$. A fundamental result is then given by the following theorem which turns out to be instrumental to obtain a topology derivation algorithm.  \newline

\begin{theorem}
\label{th:1}
Considering a wired network made by $N$ nodes, the distance between any leaf $i$ and another node $j$ can be found by applying \eqref{eq:dist}. The result is the correct value with probability 1 either for $d_1$ or $d_2$, if and only if:
	\begin{enumerate}
		\item the admittance $Y_2$ used in \eqref{eq:dist} is the one measured at node $j$, to which node $i$ is directly connected;
		\item the actual length of the line connecting $i$ and $j$ is less than $\lambda/4$, where $\lambda$ is the wavelength used to perform the admittance measurements.
	\end{enumerate}
\end{theorem}
\vspace{5pt}
\begin{corollary}
\label{cor:1}
When no parameter or measurement error exists, the topology (graph and branch lengths) of any line network in which 
	\begin{equation*}
		\max_{i,j \in N} d_{i,j} \le \frac{\lambda}{4},
	\end{equation*}
can be exactly derived by means of a recursive use of \eqref{eq:dist}, with the exploitation of the measured network admittances and the available cable parameters and loads.
\end{corollary}
\vspace{17pt}

\textit{Proof:}

A sufficient condition to obtain a unique and real solution for the distance is when the pair of nodes are a leaf and the adjacent node. An this is obvious from the physical construction of the problem and associated TL equations. 

To prove the necessity, i.e., that there exists a unique $Y_2$ for which the solution $d$ to \eqref{eq:dist} is real, we follow a probabilistic reasoning. Firstly, let's consider the plane where the impedance $Y_2$ can possibly lay and let's define an arbitrary value of it with $y_2$. Then, we note that the locus of points for which the imaginary part of $d$ is zero is a line, because $d$ is the logarithm of a polynomial function. Since any complex polynomial is holomorphic \cite{9781139171915}, it cannot be locally constant, so its imaginary part can assume one value only along a line, or a sequence of lines (and the logarithm does not influence the function in this sense).

\begin{figure}
	\centering
		\includegraphics[width=0.40\textwidth]{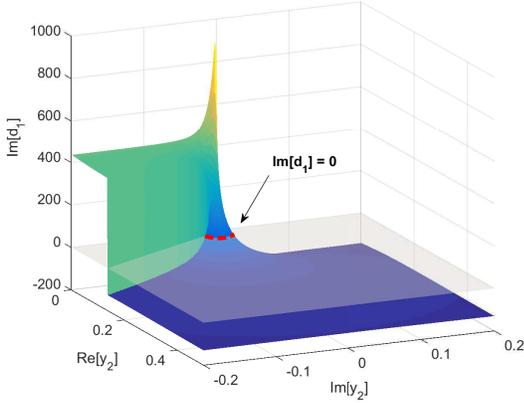}
	\caption{Example of one possible $\Im(d_1)$ as a function of the sole $y_2$. }
	\label{fig:d1imag}
\end{figure}

As an example, $\Im(d_1)$ is plotted as a function of $y_2$ in Fig. \ref{fig:d1imag}. The bold-dashed line in this figure highlights the locus of points where the imaginary part of the distances is zero. 
Secondly, let's assume to randomly pick $y_2$, i.e., the real and imaginary parts of it are independent, continuous random variables. Then, the probability that $y_2$ lays on a line is zero and consequently the probability that $\Im(d_2) = 0$ or $\Im(d_1) = 0$ is also zero. 

We can therefore state that with probability 1, the only case for which $d_1$ or $d_2$ are real valued is the case corresponding to the sufficient condition, i.e., when $Y_1$ and $Y_2$ are the network admittances of the leaf and the adjacent node. Furthermore, only one among $d_1$ and $d_2$ will be real valued. This is because the problem of finding where $\Im(d_1(y_2)) = \Im(d_2(y_2)) = 0$ implies to find the intersection between two lines. Observing the form of \eqref{eq:dist}, the number of intersection points is finite, so that $P[\Im(d_2(y_2))=\Im(d_1(y_2))= 0] = 0$. 

Finally, the assumption that the distance among the two nodes is less or equal to $\lambda/4$ is a prerequisite to invert equation \eqref{eq:two_loads} since, as we have already explained, that equation is bijective only when $d < \lambda/4$.

The corollary is an immediate consequence of the theorem. It will be constructively used in an algorithm to derive the topology graph and branch lengths in the next section.
 
 
\section{Admittance noise}
\label{sec:noise}

In real scenarios admittance measurements are perturbed by noise. In this section, a mathematical derivation of the admittance noise is performed. This noise will be related to the signal-to-noise ratio (SNR) normally defined in communications.

\begin{figure}
	\centering
		\includegraphics[width=0.25\textwidth]{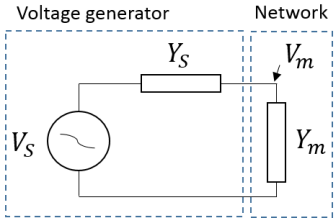}
	\caption{Sketch of a voltage measurement device. $V_S$ and $Y_S$ are the equivalent voltage generator and admittance, $Y_m$ is the network admittance at node $m$ to be measured.}
	\label{fig:partitore}
\end{figure}

An admittance meter can be represented as its Thevenin equivalent, being $V_S$ the equivalent generated voltage phasor and $Y_S$ the equivalent output admittance. When an admittance measurement is performed, then a voltage divider is created between $Y_S$ and $Y_m$, the unknown network admittance at node $m$ (see Fig.~\ref{fig:partitore}). Using the voltage divider equation, $Y_m$ can be derived as $Y_{m} = Y_SV_S/V_m-Y_S$, where $V_m$ is the phasor of the voltage drop across $Y_m$. $V_m$ is affected by the background noise present in the network, so that it can be written as $V_m = V_{m_0} + V_{m_N}$, where $V_{m_N} \sim \mathcal{CN} (0,\sigma^2_N)$ and $V_{m_0} = \mathbb{E}\left[V_m\right]$. $\mathcal{CN} (0,\sigma^2_N)$ denotes a complex Gaussian variable with zero mean and variance $\sigma^2$; $\mathbb{E}[\cdot{}]$ denotes the expectation operator. The real and imaginary noise are assumed to be independent and with the same variance $\sigma^2_N/2$. 

If we assume $V_S$ and $Y_S$ to be ideal, the noisy load admittance $Y_m$ can be written as
\begin{align}
	Y_m &= \frac{Y_S\left(V_S - V_{m_0} - V_{m_N}\right)}{V_{m_0} + V_{m_N}} \\
			&= \frac{Y_S\left(V_S - V_{m_0}\right)}{V_{m_0} + V_{m_N}} - \frac{Y_S V_{m_N}}{V_{m_0} + V_{m_N}}.
	\label{eq:ylnoise}
\end{align}
In particular, if the SNR at the node $m$, i.e. $\mathbb{E}\left[\left|V_{m_0}\right|^2\right]/\mathbb{E}\left[\left|V_{m_N}\right|^2\right]$, is sufficiently high, then \eqref{eq:ylnoise} can be simplified as
\begin{equation}
	Y_m = \frac{Y_S\left(V_S - V_{m_0}\right)}{V_{m_0}} - \frac{Y_S V_{m_N}}{V_{m_0}} = Y_{m_0} + Y_{m_N},
	\label{eq:ylnoise_sim}
\end{equation}
thus $Y_m$ can also be considered as a perfect measurement $Y_{m_0}$ corrupted by the Gaussian noise $Y_{m_N}$. To experimentally prove it, we generated thousands of realizations of $Y_m$ with different parameters and SNRs. By applying the standard Kolmogorow-Smirnow Test \cite{daniel1990applied}, we discovered that the simplification introduced in \eqref{eq:ylnoise_sim} is valid when SNR $>$ 35~dB. Such a value of SNR is easily exceeded for example when the measure is done according to the PLC standards, where in the worst condition at few kHz the background noise can reach -70~dBm, while the corresponding transmit power is around -15~dBm \cite{7037264}.

When \eqref{eq:ylnoise_sim} holds, we can define the Admittance to Noise Ratio (ANR) as 
\begin{equation}
	ANR = \frac{\mathbb{E}\left[\left|Y_{m_0}\right|^2\right]}{\mathbb{E}\left[\left|Y_{m_N}\right|^2\right]} 
			\simeq \frac{\left|Y_{m_0}\right|^2}{\sigma^2_{N}}  = \frac{\left|V_S-V_{m_0}\right|^2}{\sigma^2_{N}},
	\label{eq:ANR}
\end{equation}
where $\sigma^2_N$ is the variance of $V_{m_N}$ and the second equivalence holds when $Y_{m_0}$ can be considered static over time, i.e. within the coherence time of the channel. 
\begin{figure}
	\centering
		\includegraphics[width=0.43\textwidth]{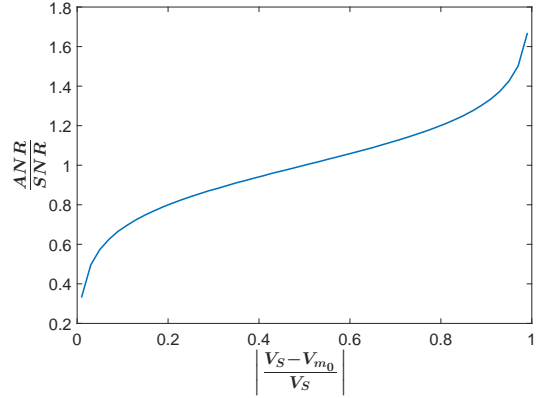}
	\caption{Normalized plot of \eqref{eq:ANR} for different values of $V_{m_0}$}
	\label{fig:anr_vs_vl}
\end{figure}
As we can see in Fig.~\ref{fig:anr_vs_vl}, when $V_{m_0}$ is close to $V_S$, the admittance noise is amplified, leading to an ANR lower than the SNR. Vice versa, when a high value of ANR is wanted, then $V_{m_0}$ has to be as little as possible. A little value of $V_{m_0}$ can be easily achieved by using $|Y_S| \ll |Y_{m_0}|$.

\section{Topology Derivation}
\label{sec:topdev}

In this section, we present an algorithm that relies on Theorem \ref{th:1} and that allows to derive the topology of a general tree-structured wired network, together with the length of all the lines connecting the nodes. 

Let $\mathcal{T}=(\mathcal{N},\mathcal{L})$ denote the topology of a network, where $\mathcal{N}$ is the set of all the $N$ nodes of the network and $\mathcal{L}$ is the set made of all the physical connections between two network elements. The terminal nodes, i.e. those nodes that are connected to the rest of the network with a single branch, are referred to as leafs.  

To each node of the network $i \in \mathcal{N}$ we associate a load admittance $Y_{L_i}$, that is characteristic of the device plugged to the network, and a network admittance $Y_{N_i}$ that is the admittance measured at node $i$ comprising the load at that node. The cable parameters for each line $l \in \mathcal{L}$ departing from node $i$ are assumed to be known. Such cable parameters are the propagation constant $\Gamma_l$ and the characteristic admittance of the line $Y_{C_l}$. Clearly, in a uniform network all cables have identical parameters.

Algorithm \ref{alg:mine} offers a method to derive $\mathcal{T}$ taking as inputs the parameters $Y_{L}$, $\Gamma$, $Y_{C}$, $Y_{N}$, that are known for each node and branch of the network. The last parameter needed is the ANR that is assumed to be the same for every node and can be sensed by the modem during the calibration period of each data transmission. The core idea of the algorithm is that if the imaginary part of one of the two computed cable lengths is small enough, then also the error on the real part is small, and the two nodes considered are with high probability connected.

The algorithm firstly considers the full set of nodes $\mathcal{N}$, and it assumes them to be, potentially, leafs. With this assumption, \eqref{eq:dist} is applied to every pair of nodes $i$ and $k$. If the result provides a solution whose imaginary part is greater than a certain threshold, then node $i$ cannot be a leaf and $k$ cannot be a directly connected node. If instead the imaginary part of one solution is lower than the threshold, the algorithm states that $i$ is a leaf and the two nodes $i$ and $k$ are directly connected with a branch having real length $\Re[d]$. 
After having found all actual leafs, their load admittances are carried back to the connected internal nodes, to form a reduced network. The information about the connections and line lengths is contextually stored. The algorithm can also detect false positives and false negatives; in case of detection, the algorithm is interrupted. 

The procedure described in the previous paragraph iterates until the whole network is reduced to a single node. The final outputs are the complete topology $\mathcal{T}=(\mathcal{N},\mathcal{L})$ and the complete set of lengths for the $\mathcal{L}$ links. 

\begin{algorithm}
\caption{Topology derivation}
\label{alg:mine}
\begin{algorithmic}[1]
	\Require	$Y_m$ and $Y_L$ for each node, $\Gamma$ and $Y_C$ for each cable.
	\Statex
\Procedure {$[\textbf{d},\mathcal{T}]$ = Derivation(${\bf y}_L$,${\bf y}_m$,${\bf y}_C$,$\boldsymbol\gamma$,ANR)}{}
\State $\textbf{gl} \gets \text{1: length of $\textbf{y}_L$}$
\State $\text{thr} \gets \text{f(ANR)}$
\While{$\text{length of ($\textbf{gl}$)} > 1$}
	\For{$i \in \textbf{gl}$}
		\State $j = 0$
		\For{$k \in (\textbf{gl} \setminus i)$}
		\State $\textbf{x}_{(\textbf{gl}(i),\textbf{gl}(k))} \gets \text{equations \ref{eq:dist}}$
		\If{$(\Im\text[d] \subset \textbf{x}_{(\textbf{gl}(i),\textbf{gl}(k))}) < \text{thr}$}
			\State $j \gets k$
		\EndIf
	\EndFor
	\State $\textbf{d}_{(\textbf{gl}(i),\textbf{gl}(j))} \gets \textit{d}$
	\State $\mathcal{N} \gets \mathcal{N} \cup \textbf{gl}(i)$
	\State $\mathcal{L} \gets \mathcal{L} \cup (\textbf{gl}(i),\textbf{gl}(j))$
	\State ${\bf y}_L (j) \gets {\bf y}_L (j) + \textit{carryback (${\bf y}_L (i), d$)}$
	\EndFor
\State $\textbf{gl} \gets \textbf{gl}\setminus\textbf{gl}(i)$
\EndWhile
\EndProcedure
\Statex
\Ensure Network topology and length of all the branches.
\end{algorithmic}
\end{algorithm}

\section{Results}
\label{sec:results}
In this section we firstly introduce the physical layer simulator that we developed in order to test Algorithm \ref{alg:mine}; then we present and comment the results, and finally we discuss the open issues.

\begin{figure}
	\centering
	\includegraphics[width=0.46\textwidth]{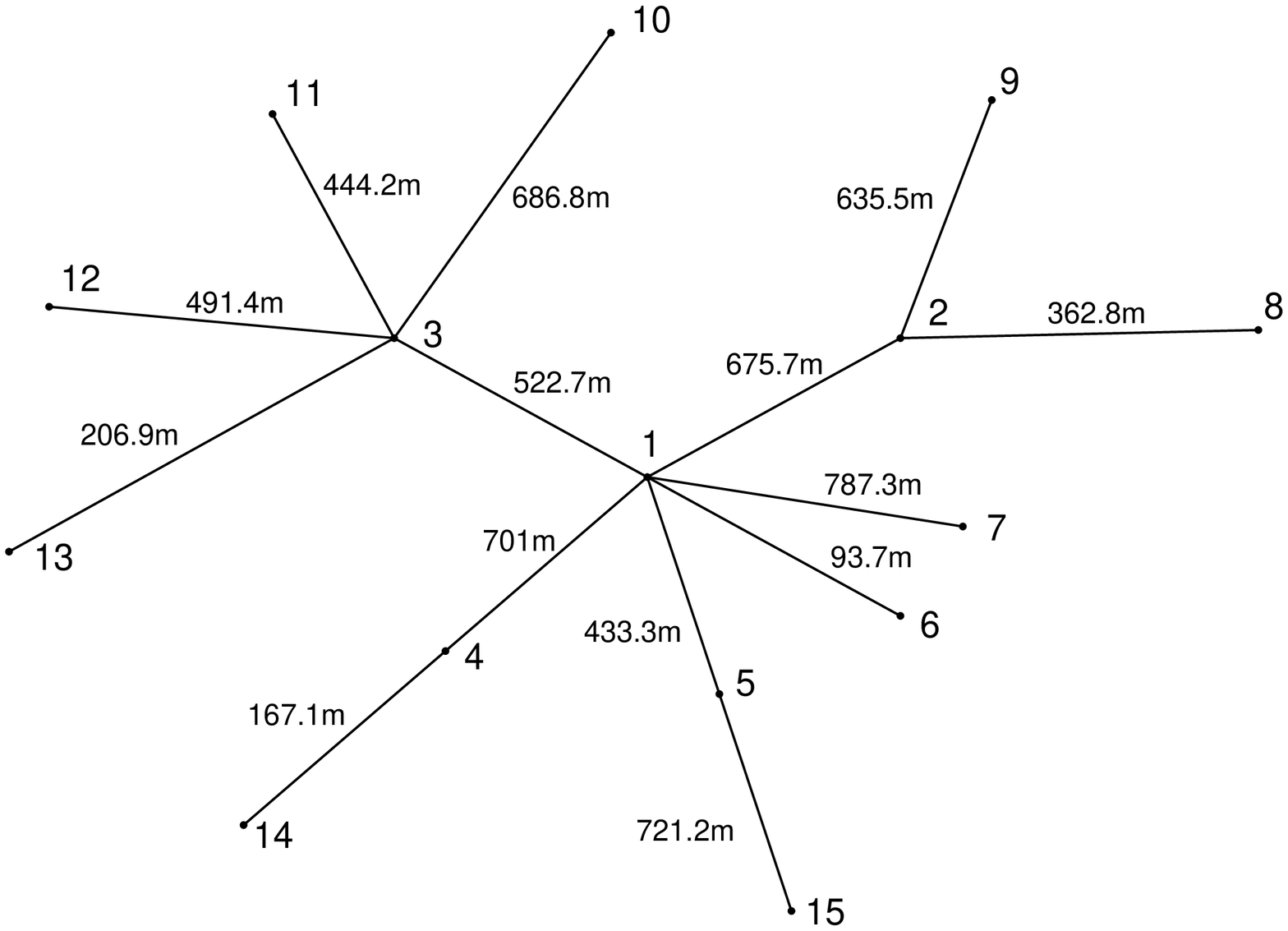}
	\caption{Sketch of a network realized by the simulator.}
	\label{fig:example_topology}
\end{figure}

To test Algorithm \ref{alg:mine}, we developed a network simulator that creates a random tree network (see Fig. \ref{fig:example_topology}) and computes the network admittance at each node. The network simulator is based on the TL theory and exploits the concentrated parameter model used in \cite[Sec. III.C]{tonello2011bottomup} to describe the cable parameters. 


Then, complex Gaussian noise is added to each network admittance according to the ANR specified by the user. Algorithm \ref{alg:mine} is finally applied to derive the topology. The simulator outputs whether a topology has been found or not, and in the positive case it outputs also $[\textbf{d},\mathcal{T}]$.
As an empirical proof of Theorem \ref{th:1} and Corollary \ref{cor:1}, we found that when no noise is added to the measurements, $[\textbf{d},\mathcal{T}]$ is correctly derived in 100$\%$ of the cases, independently from the size of the network and the number of nodes.

\subsection{Results with background noise}
\begin{figure}
	\centering
	\subfloat[][$f_0$ = 10~kHz]
		{\includegraphics[width=0.24\textwidth]{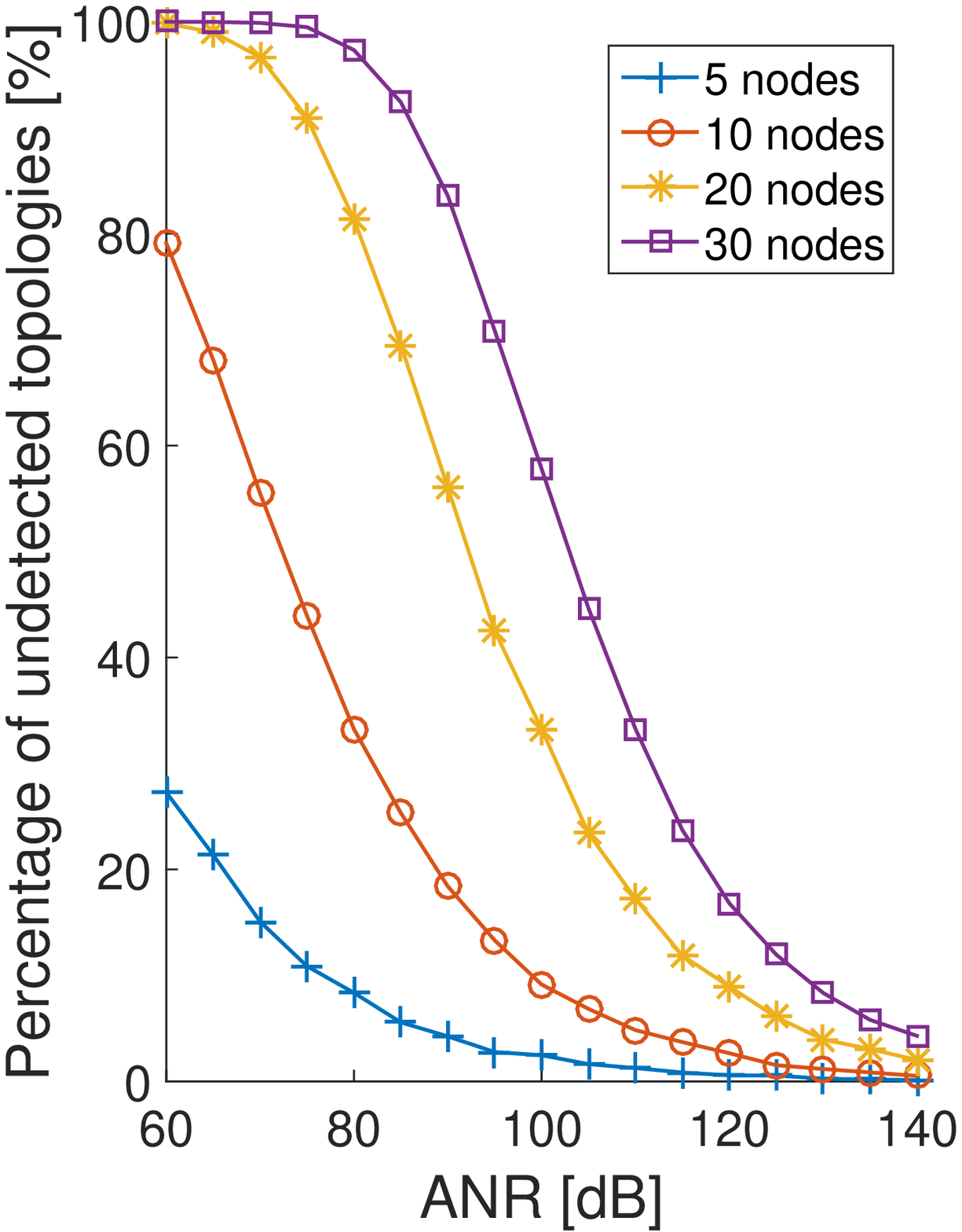}}
	\subfloat[][$f_0$ = 30~kHz]
		{\includegraphics[width=0.24\textwidth]{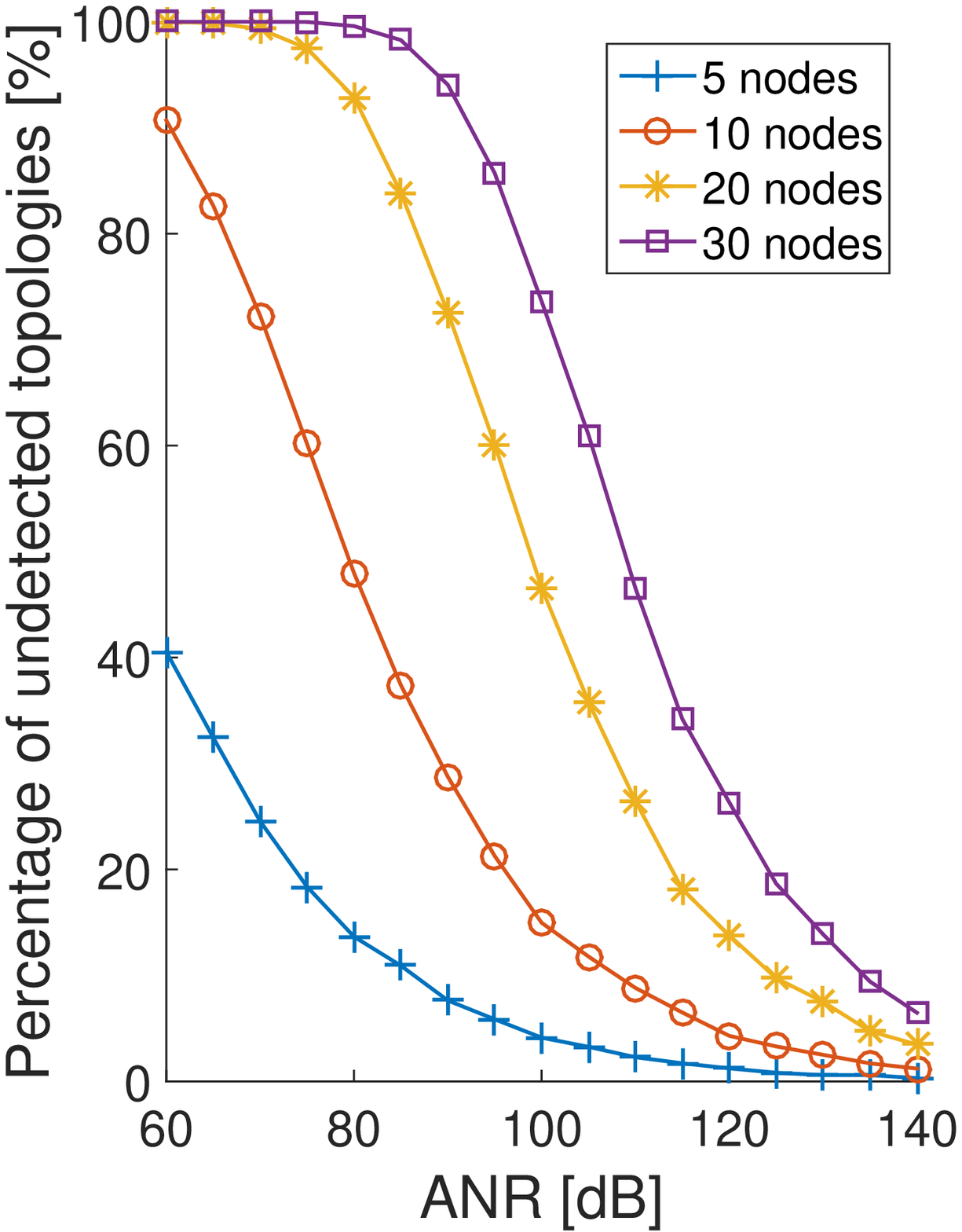}}	
	\caption{Percentage of correctly derived topologies as function of the ANR. Test run with a maximum cable length of 1.4~km. }
	\label{fig:thresholding_effect_all_fit}
\end{figure}

As explained in Section \ref{sec:noise}, the network background noise causes the network admittance measurements to be affected by error. This noise deteriorates the performance of Algorithm \ref{alg:mine}, as shown in Fig.~\ref{fig:thresholding_effect_all_fit}. For the depicted test, an adaptive threshold has been used in order to get the best performance for each ANR. The results confirm that the performance of the algorithm decreases with the increasing number of nodes. Moreover lower measurement frequencies give better results: in fact, using a noisy $Y_2$ in \eqref{eq:two_loads}, we see that the error in $Y_1$ grows as a function of the measurement frequency and the length of the cable. We remark that if a topology is correctly identified, it means that also all the branch lengths are identified with a negligible error. In fact, a consistent error in the computation of a single branch length would deeply affect the subsequent iterations of Algorithm \ref{alg:mine}, thus leading to a topology identification error.

\begin{figure}
	\centering
	\subfloat[][$f_0$ = 10~kHz]
		{\includegraphics[width=0.24\textwidth]{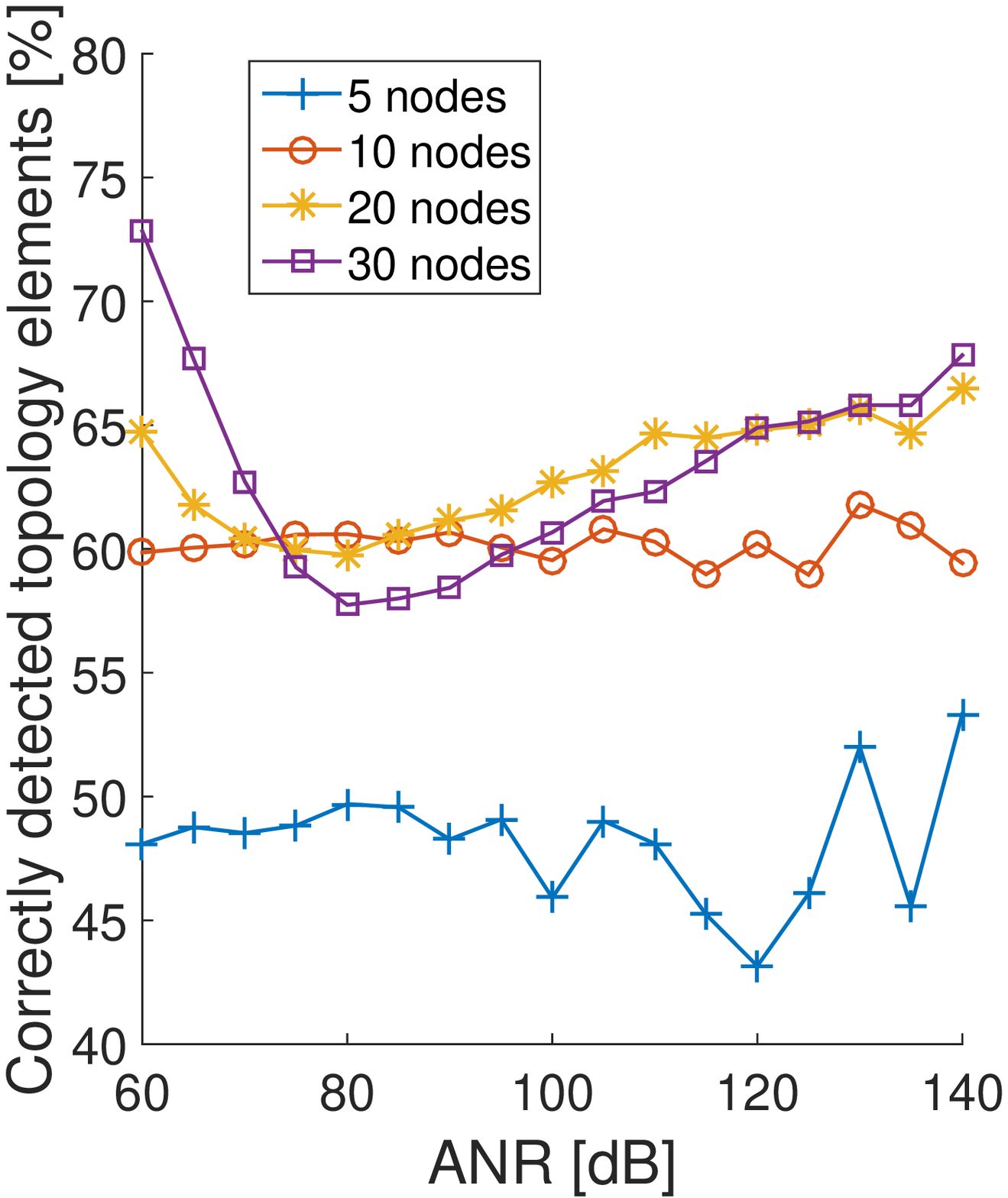}}
	\subfloat[][$f_0$ = 30~kHz]
		{\includegraphics[width=0.24\textwidth]{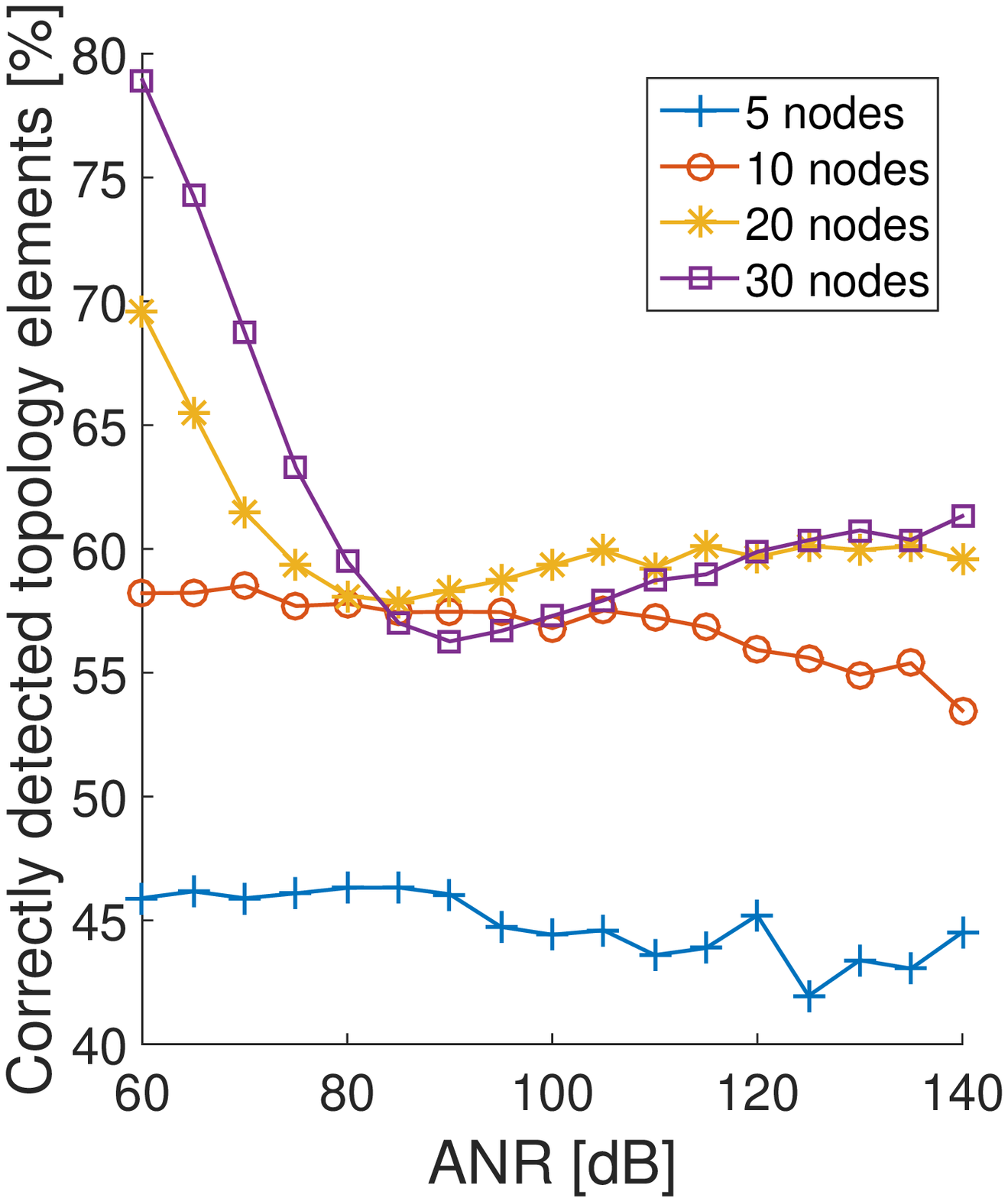}}	
	\caption{Percentage of correctly detected junctions when the topology has not been completely derived. The same test conditions of Fig.\ref{fig:thresholding_effect_all_fit} are used.}
	\label{fig:correctly_detected_when_bad}
\end{figure}

Fig.~\ref{fig:correctly_detected_when_bad} shows the percentage of correctly detected topology elements, i.e. branches, when the topology is not completely derived. Since the full topology is finally not derived, we infer that correctly detected topology elements of Fig.~\ref{fig:correctly_detected_when_bad} are affected by an error in the branch length derivation that is not negligible. However two aspects arise: the percentage does not change with the ANR, and it increases with the number of nodes. Both these aspects are a consequence of the fact that when the ANR decreases, also the number of correctly derived topologies decreases, so more topologies are taken into account for the computation here considered. The steep increments at low ANR for 20 and 30 nodes are due to the fact that for low ANRs almost no topology is completely derived. Nevertheless, Fig.~\ref{fig:correctly_detected_when_bad} shows that a consistent part of the topological information of the network can be retrieved even when the topology is not completely derived.

Table~\ref{tab:BondsOfTheValidity} provides some information about typical values of ANR that we could encounter by operating in the PLC band and fulfilling PLC norms. According to Table~\ref{tab:BondsOfTheValidity}, if the measurement is performed at $f_0 =$ 10~kHz, the ANR is approximately 100~dB, so that from Fig.~\ref{fig:thresholding_effect_all_fit} we see that for 10 nodes more than 90\% of the topologies is correctly detected. Moreover, about 60\% of the connections within the remaining topologies is correctly identified (see Fig.~\ref{fig:correctly_detected_when_bad}). 

\begin{table*}[t]
	\centering
	\caption{Typical parameter values used to evaluate the proposed algorithm.}
		\begin{tabular}{l|l|l|l}
		\toprule
		& Cenelec (3--150~kHz) & FCC (150--500~kHz) & Broad-Band (2--30~MHz) \\
		\midrule
		Maximum cable length & 16.6~km & 330~m & 25~m \\
		Average Ground noise \cite{7037264} & -70 -- -90~dBm/Hz & -90 -- -110~dBm/Hz & $<$ -110~dBm/Hz \\
		Transmitted power & $\sim$-15~dBm/Hz & -22~dBm/Hz & -55~dBm/Hz \\
		ANR & 99 -- 135~dB & 122 -- 158~dB & $>$99~dB \\
		\bottomrule
			
		\end{tabular}
		\vspace{0.2cm}
	\caption*{Remarks: the maximum cable length depends on its propagation constant (herein light velocity $v_c$ = 2e8~m/s). The data about the transmitted power has been retrieved from \cite{lampe2016power} and from the standard IEEE 1901.2a-2015. As for the ANR, here we consider the condition in which the measured network voltage has a small absolute value compared to the voltage provided by the voltmeter, so that ANR = 1.8 SNR (see Fig.~\ref{fig:anr_vs_vl}).}
	\label{tab:BondsOfTheValidity}
	\vspace{-0.5cm}
\end{table*}

\section{Remarks and Open Issues}
\label{sec:remarks}
The analysis reported so far has shown that admittance measurements can be exploited to gain knowledge about the topology of a wired network. In particular, admittance measurements can be used to provide a real-time solution that can partially or completely derive the network topology, as well as the length of the branches connecting the nodes, all in a unified algorithm. The algorithm in presence of noise has been tested using typical values from PLC in the context of SMG.

The approach that we have discussed opens further questions on some open issues as we discuss in the following.

\begin{enumerate}

\item \label{item1}The frequency at which measurement have to be made, depends on the node distances and cannot be freely chosen. However, this is not very restrictive.
 
\item \label{item2}The speed of the measurement is also important since the status of the network may change over time. In fact, although the length and the node connections can be considered time-invariant, the loads may change. Therefore, the coherence time of the topology must be larger than the measurement time. 
 
\item \label{item3}Synchronization and coordination of the admittance measurements can also be required in a network with a time variant status.  
This could be done for example with a GPS coordination system or with communication modules.

\item \label{item4}An important aspect is the presence of measurement errors and uncertainty in the required parameters, i.e., loads and cable parameters. Proper estimation techniques can be used to tackle such a problem and also to enhance the performance of Algorithm \ref{alg:mine} in the presence of noise. Interestingly, the proposed topology derivation technique may be used inversely to track the cables deterioration, by sensing the increment of the error in the derivation of the branch lengths over time.

	
\end{enumerate}

We point out that the open issues \ref{item2} and \ref{item3} mentioned in this section have not yet been fully considered in the literature of wired network topology estimation, while open issue \ref{item4} is shared with \cite{5256192}. All these open issues provide stimulus for further research endeavors.

\section{Conclusion}

In this paper, we have addressed the question: ``Can we exploit admittance measurements to derive the topology of a wired network ?''
The approach differs from others presented in the literature, which use reflectometry or ToA estimation followed by topology inference algorithms. We have shown that the admittance measurement based approach can allow the derivation of the topology and the length of all branches by performing admittance measurements at all nodes.  It is the direct application of the derived Theorem 1, which states that it is possible to identify whether a pair of nodes in a network are respectively a leaf and a directly connected node, and if so, a solution to the derivation of the length of the line connecting them can be found. We have further shown that the admittance measurements are perturbed by complex Gaussian noise in the presence of line background noise when the SNR at the receiver is greater than 35~dB. Moreover, the ANR is greater than the SNR when the internal impedance of the measurement device is greater than the measured impedance. 
These findings have been used to develop an algorithm that derives the topology of a wired network provided that admittance measurements are done at all the network nodes and that the cable parameters and the loads are known. Future research directions in this topic have also been discussed and include: robust topology derivation in the presence of parameter errors; application of the admittance based topology derivation algorithm to track the cables deterioration.


\bibliographystyle{IEEEtran}
\bibliography{IEEEabrv,femtocell_biblio}

\begin{thebibliography}{10}
\providecommand{\url}[1]{#1}
\csname url@samestyle\endcsname
\providecommand{\newblock}{\relax}
\providecommand{\bibinfo}[2]{#2}
\providecommand{\BIBentrySTDinterwordspacing}{\spaceskip=0pt\relax}
\providecommand{\BIBentryALTinterwordstretchfactor}{4}
\providecommand{\BIBentryALTinterwordspacing}{\spaceskip=\fontdimen2\font plus
\BIBentryALTinterwordstretchfactor\fontdimen3\font minus
  \fontdimen4\font\relax}
\providecommand{\BIBforeignlanguage}[2]{{%
\expandafter\ifx\csname l@#1\endcsname\relax
\typeout{** WARNING: IEEEtran.bst: No hyphenation pattern has been}%
\typeout{** loaded for the language `#1'. Using the pattern for}%
\typeout{** the default language instead.}%
\else
\language=\csname l@#1\endcsname
\fi
#2}}
\providecommand{\BIBdecl}{\relax}
\BIBdecl

\bibitem{lampe2016power}
L.~Lampe, A.~M. Tonello, and T.~G. Swart, Eds., \emph{Power Line
  Communications: Principles, Standards and Applications from Multimedia to
  Smart Grid}.\hskip 1em plus 0.5em minus 0.4em\relax Wiley, 2016.

\bibitem{gungor2011sgt}
V.~Gungor, D.~Sahin, T.~Kocak, S.~Ergut, C.~Buccella, C.~Cecati, and G.~Hancke,
  ``Smart grid technologies: Communication technologies and standards,''
  \emph{IEEE Transactions on Industrial Informatics}, vol.~7, no.~4, pp.
  529--539, Nov 2011.

\bibitem{1007375}
S.~Galli and D.~L. Waring, ``Loop makeup identification via single ended
  testing: beyond mere loop qualification,'' \emph{IEEE Journal on Selected
  Areas in Communications}, vol.~20, no.~5, pp. 923--935, Jun 2002.

\bibitem{5256192}
C.~Sales, R.~M. Rodrigues, F.~Lindqvist, J.~Costa, A.~Klautau, K.~Ericson,
  J.~R. i~Riu, and P.~O. Borjesson, ``Line topology identification using
  multiobjective evolutionary computation,'' \emph{IEEE Transactions on
  Instrumentation and Measurement}, vol.~59, no.~3, pp. 715--729, March 2010.

\bibitem{5953512}
R.~M. Rodrigues, C.~Sales, A.~Klautau, K.~Ericson, and J.~Costa, ``Transfer
  function estimation of telephone lines from input impedance measurements,''
  \emph{IEEE Transactions on Instrumentation and Measurement}, vol.~61, no.~1,
  pp. 43--54, Jan 2012.

\bibitem{1430477}
G.~Bumiller, L.~Lu, and Y.~Song, ``Analytic performance comparison of routing
  protocols in master-slave plc networks,'' in \emph{International Symposium on
  Power Line Communications and Its Applications, 2005.}, April 2005, pp.
  116--120.

\bibitem{ahmed2012topology2}
M.~Ahmed and L.~Lampe, ``Parametric and nonparametric methods for power line
  network topology inference,'' in \emph{Power Line Communications and Its
  Applications (ISPLC), 2012 16th IEEE International Symposium on}, March 2012,
  pp. 274--279.

\bibitem{7431885}
C.~Zhang, X.~Zhu, Y.~Huang, and G.~Liu, ``High-resolution and low-complexity
  dynamic topology estimation for {PLC} networks assisted by impulsive noise
  source detection,'' \emph{IET Communications}, vol.~10, no.~4, pp. 443--451,
  2016.

\bibitem{erseghe2013topology}
T.~Erseghe, S.~Tomasin, and A.~Vigato, ``Topology estimation for smart micro
  grids via powerline communications,'' \emph{Signal Processing, IEEE
  Transactions on}, vol.~61, no.~13, pp. 3368--3377, July 2013.

\bibitem{lampe2013tomography}
M.~O. Ahmed and L.~Lampe, ``Power line communications for low-voltage power
  grid tomography,'' \emph{IEEE Transactions on Communications}, vol.~61,
  no.~12, pp. 5163--5175, Dec. 2013.

\bibitem{6338332}
A.~Milioudis, G.~Andreou, and D.~Labridis, ``Enhanced protection scheme for
  smart grids using power line communications techniques -- part {I}: Detection
  of high impedance fault occurrence,'' \emph{Smart Grid, IEEE Transactions
  on}, vol.~3, no.~4, pp. 1621--1630, Dec 2012.

\bibitem{6954542}
------, ``Detection and location of high impedance faults in multiconductor
  overhead distribution lines using power line communication devices,''
  \emph{Smart Grid, IEEE Transactions on}, vol.~6, no.~2, pp. 894--902, March
  2015.

\bibitem{Pozar}
D.~M. Pozar, \emph{Microwave Engineering -- Fourth Edition}.\hskip 1em plus
  0.5em minus 0.4em\relax John Wiley \& Sons, 2011.

\bibitem{tonello2011bottomup}
A.~Tonello and F.~Versolatto, ``Bottom-up statistical {PLC} channel modelling -
  part \uppercase{I}: Random topology model and efficient transfer function
  computation,'' \emph{IEEE Transactions on Power Delivery}, vol.~26, no.~2,
  pp. 891--898, Apr. 2011.

\bibitem{lampe2013tomography2}
L.~Lampe and M.~Ahmed, ``Power grid topology inference using power line
  communications,'' in \emph{Smart Grid Communications (SmartGridComm), 2013
  IEEE International Conference on}, Oct 2013, pp. 336--341.

\bibitem{9781139171915}
\BIBentryALTinterwordspacing
G.~A. Jones and D.~Singerman, \emph{Complex Functions}.\hskip 1em plus 0.5em
  minus 0.4em\relax Cambridge University Press, 1987, cambridge Books Online.
  [Online]. Available: \url{http://dx.doi.org/10.1017/CBO9781139171915}
\BIBentrySTDinterwordspacing

\bibitem{daniel1990applied}
\BIBentryALTinterwordspacing
W.~Daniel, \emph{Applied nonparametric statistics}, ser. The Duxbury advanced
  series in statistics and decision sciences.\hskip 1em plus 0.5em minus
  0.4em\relax PWS-Kent Publ., 1990. [Online]. Available:
  \url{https://books.google.at/books?id=0hPvAAAAMAAJ}
\BIBentrySTDinterwordspacing

\bibitem{7037264}
M.~Girotto and A.~M. Tonello, ``Improved spectrum agility in narrow-band {PLC}
  with cyclic block {FMT} modulation,'' in \emph{2014 IEEE Global
  Communications Conference}, Dec 2014, pp. 2995--3000.

\end{thebibliography}
\end{document}